\documentclass[twocolumn,prl,aps,longbibliography,superscriptaddress]{revtex4-1}

\usepackage{graphicx}
\usepackage{amsmath}
\usepackage{amsfonts}
\usepackage{amssymb}
\usepackage{xcolor}
\usepackage{hyperref}
\hypersetup{colorlinks=true,allcolors=blue}
\usepackage[T1]{fontenc}

\usepackage{footnote}
\usepackage{textcomp}

\renewcommand{\t}[1]{\textrm{#1}}
\newcommand{\nn}{\nonumber\\}

\newcommand{\g}{\gamma}

\renewcommand{\r}{\rho}
\newcommand{\s}{\sigma}

\newcommand{\B}{\mathcal{B}}
\newcommand{\I}{\mathcal{I}}

\newcommand{\N}{\mathcal{N}}

\renewcommand{\P}{\mathcal{P}}
\newcommand{\Q}{\mathcal{Q}}

\newcommand{\+}{^\dagger}
\renewcommand{\>}{\rangle}
\newcommand{\<}{\langle}

\begin{document}

%\title{Photoemission of a quantum dot: non-Markovianity and the quantum regression theorem}

%Alternative title suggestions:
%\title{Testing the quantum regression theorem for photon emission from a quantum dot}
%\title{Limits of the quantum regression theorem for photon emission from a quantum dot}
%\title{Quantifying photon emission properties of a quantum dot beyond the quantum regression theorem}
%\title{Beyond the quantum regression theorem}
\title{Accuracy of the quantum regression theorem for photon emission from a quantum dot}

\author{M. Cosacchi}
%\email{michael.cosacchi@uni-bayreuth.de}
\affiliation{Theoretische Physik III, Universit{\"a}t Bayreuth, 95440 Bayreuth, Germany}
\author{T. Seidelmann}
\affiliation{Theoretische Physik III, Universit{\"a}t Bayreuth, 95440 Bayreuth, Germany}
\author{M. Cygorek}
\affiliation{Heriot-Watt University, Edinburgh EH14 4AS, United Kingdom}
\author{A. Vagov}
\affiliation{Theoretische Physik III, Universit{\"a}t Bayreuth, 95440 Bayreuth, Germany}
\affiliation{ITMO University, St. Petersburg, 197101, Russia}
\author{D. E. Reiter}
\affiliation{Institut f\"ur Festk\"orpertheorie, Universit\"at M\"unster, 48149 M\"unster, Germany}
\author{V. M. Axt}
\affiliation{Theoretische Physik III, Universit{\"a}t Bayreuth, 95440 Bayreuth, Germany}

\begin{abstract}
The quantum regression theorem (QRT) is the most-widely used tool for calculating multitime correlation functions for the assessment of quantum emitters.
It is an approximate method based on a Markov assumption for the environmental coupling.
In this work we quantify properties of photons emitted from a single quantum dot coupled to phonons.
For the single-photon purity and the indistinguishability,
we compare numerically exact path-integral results with those obtained from the QRT. 
It is demonstrated that the QRT systematically overestimates the influence of the environment for typical quantum dots used in quantum information technology.
\end{abstract}
\maketitle

%\textbf{Introduction}
To be used as photon sources for quantum information technology \cite{Michler2000,Senellart2017}, semiconductor quantum dots (QDs) must deliver photons with high-quality characteristics such as a high brightness, a perfect single-photon purity, and indistinguishability.
However, due to the electron-phonon interaction in QDs these quantities can be degraded \cite{Gerhardt2018,Wang2020}.
In the current race for the perfect single-photon source \cite{Vural2020,Thomas2021} with achieved purities and indistinguishablities close to unity \cite{Ding2016,Somaschi2016,Wang2020,Arakawa2020,
Tomm2021,Thomas2021b,Zhai2021}, it is crucial to understand the influence of the phonon-induced dephasing on the properties of emitted photons.
The coupling to environmental phonons has been shown to lead to several important phenomena like the phonon sidebands \cite{Besombes2001,Krummheuer2002}, damping of Rabi oscillations \cite{Foerstner2003,Machnikowski2004, Vagov2004,Vagov2007,Ramsay2010a}, and the possibility for a dynamic decoupling of electronic and phononic subsystems \cite{kaldewey2017demonstrating,Luker2019}, or degradation of photon properties \cite{Iles-Smith2017b}.

The quantum regression theorem (QRT) known from quantum optics is probably the most-widely used standard tool to investigate the above photon properties \cite{Breuer2002}.
In essence, the QRT prescribes to calculate the two- (or multi-) time correlation functions using the same dynamical equation for both the (real-) time and the delay-time arguments, as is used to determine the time evolution of the single-time correlations.
Solving an initial value problem for the delay-time propagation as done in the QRT disregards the memory that has build up until the start of the propagation and thus the use of the QRT may become critical when used for non-Markovian dynamics.
With the help of the QRT, multitime correlation functions yielding, e.g., the purity and indistinguishability can be deduced.
The QRT can be extended such that it also accounts for the electron-phonon interaction \cite{Kaer2013,Roy-Choudhury2015,McCutcheon2016,Iles-Smith2017,Gustin2018}.
For our current study it is most important that phonons are known to induce non-Markovian dynamics \cite{Kaer2010,mccutcheon2010qua,Kaer2013,
McCutcheon2016,Carmele2019,Reiter2019} which provides a situation where the QRT may come to its limits \cite{Renger2002,Breuer2016,deVega2017,Guarnieri2014,
Kurt2021}.
Because the QRT is an approximation it is not always clear, whether the assumptions made in the derivation are fulfilled.

Testing the limits of the QRT has recently become possible by a path-integral approach to calculate multitime photon correlation functions \cite{Cosacchi2018}.
This approach is numerically exact meaning that the time-dependent solution to the many-body model Hamiltionian is obtained without any further approximations and thus the phonon influence including its non-Markovian part is fully taken into account \cite{Vagov2011,Barth2016,Cygorek2017}.
The accuracy of the result is controlled by choosing an appropriate time discretization and memory length. 

In this work, we explore the limits of the QRT approximation for calculating multitime correlation functions using a QD coupled to phonons as an example.
To compare numerically exact results with the QRT approximation in the most transparent way, we implement the QRT directly within the path-integral method.
Since apart from the QRT no further approximations are involved, this approach offers a direct way to evaluate the influence of the QRT on the multitime correlations.
Details of the implementations are found in the Supplemental Material (SM) \cite{supp}.

We demonstrate that the QRT systematically overestimates the phonon impact on the indistinguishability, in particular for standard GaAs QDs relevant for technological applications \cite{Hepp2020,Bremer2020,Holewa2020,Srocka2020,
Phillips2020,Brash2019,Scholl2020,Manna2020}.
We show that this is connected to the non-Markovian part of the dynamics.
In contrast, the QRT yields quantitatively correct results for the purity.

\begin{figure*}[t]
	\centering
	\includegraphics[width=\textwidth]{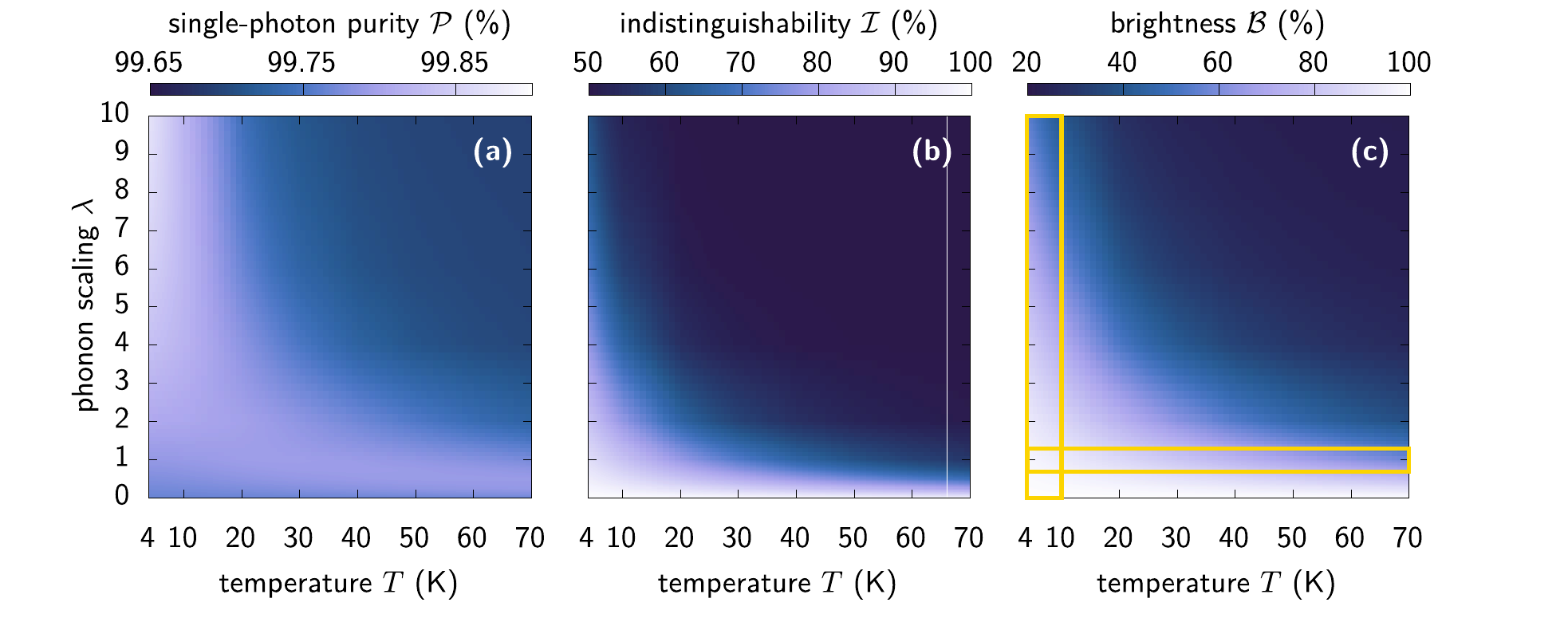}
\caption{The single-photon purity (a), the indistinguishability of two successively emitted photons (b), and their brightness (c) in a two-level QD for a temperature range between $4\,$K and $70\,$K and phonon scalings from $0$ to $10$.
Yellow rectangles in panel (c) mark the physically important parameter regime of GaAs around $\lambda=1$ for different temperatures and different phonon scalings for temperatures below $10\,$K.}
	\label{fig:I_and_B}
\end{figure*}

%\textbf{Hamiltonian}
We consider a model where a two-level QD can emit photons and interacts with environmental longitudinal acoustic (LA) phonons \cite{Krummheuer2002,Axt2005,Reiter2019}.
For the calculations we consider GaAs QDs of radius $3\,$nm and standard material parameters for the phonon coupling with the exception that we introduce a scaling factor $\lambda$ modifying the overall coupling amplitude.
Details of the model are found in the SM \cite{supp}.
We assume that this scaling is a variable in the interval $\lambda \in [0,10]$, where $0$ means absence of phonons, $1$ corresponds to the GaAs QDs and larger values allow us to explore strongly coupled QD-phonon systems \cite{Vagov2011,Luker2017}.
Larger couplings $1<\lambda\leq10$ can be found in piezoelectric materials like GaN \cite{Krummheuer2005}. We further account for the radiative decay of the QD exciton by introducing a Lindblad superoperator, setting the radiative decay rate to $\g=1\,$ns$^{-1}$. The QD is excited by an external laser pulse with a Gaussian envelope. We consider a resonant \footnote{Note that the excitation has to be resonant to the polaron shifted exciton energy, when phonons are taken into account.} excitation scheme with a $\pi$-pulse of $3$~ps length (FWHM) \footnote{For resonant excitation, this value has been shown both experimentally \cite{Ding2016} and theoretically \cite{Gustin2018} to be a favorable value concerning the photonic figures of merit.} to prepare the excited state in the QD \cite{Ding2016,Gustin2018}.
Using this model, we can then calculate the photonic properties.

%\textbf{Single-photon purity}
The single-photon purity $\P$ is defined as
\begin{align}
\mathcal{P}=1-p\qquad\text{with} \qquad 
p=\frac{\int_{-T_{\t{Pulse}}/2}^{T_{\t{Pulse}}/2}d\tau\,G^{(2)}(\tau)}{\int_{T_{\t{Pulse}}/2}^{3 T_{\t{Pulse}}/2}d\tau\,G^{(2)}(\tau)}\, .
\end{align}
$T_{\t{Pulse}}$ is the separation of the pulses in the excitation pulse train and
\begin{subequations}
\begin{align}
\label{eq:G2_tau}
G^{(2)}(\tau):=\,&\lim_{T\to\infty}\frac{1}{T}\int_{-T}^T dt\, G^{(2)}(t,\tau)\, ,\\
\label{eq:G2_t_tau}
G^{(2)}(t,\tau):=\,&\<\s_X\+(t)\s_X\+(t+\tau)\s_X(t+\tau)\s_X(t)\>
\end{align}
\end{subequations}
with $\s_X$ describing the QD transition from the excited to the ground state.
$\P$ is a measure for the single-photon component of the photonic state \cite{Michler2000,Santori2001,Santori2002,He2013,Somaschi2016,Wei2014Det,
Ding2016,Schweickert2018}.
It is measured using a Hanbury Brown-Twiss setup \cite{Hanbury1956}, which is a coincidence measurement and can thus be modeled with a second-order two-time correlation function $G^{(2)}(\tau)$. $\P=1$ implies a perfect single-photon purity. The quantity has no lower bound, $-\infty<\P\leq1$, since $p$ can be larger than one in the case of bunching instead of antibunching behavior.

%\textbf{Indistinguishability}
The indistinguishability $\I$ of two successively emitted photons is obtained as
\begin{align}
\mathcal{I}=1-p_{\t{HOM}} \quad \text{with}\quad 
p_{\t{HOM}}=\frac{\int_{-T_{\t{Pulse}}/2}^{T_{\t{Pulse}}/2}d\tau\,G^{(2)}_{\t{HOM}}(\tau)}{\int_{T_{\t{Pulse}}/2}^{3 T_{\t{Pulse}}/2}d\tau\,G^{(2)}_{\t{HOM}}(\tau)}
\end{align}
with the correlation functions  \cite{Kiraz2004,Fischer2016,Gustin2018}
\begin{subequations}
\begin{align}
\label{eq:G2HOM_tau}
G^{(2)}_{\t{HOM}}(\tau):=\,&\lim_{T\to\infty}\frac{1}{T}\int_{-T}^T dt\, G^{(2)}_{\t{HOM}}(t,\tau)\\
\label{eq:G2HOM_t_tau}
G^{(2)}_{\t{HOM}}(t,\tau):=\,&\frac{1}{2}\big[
\<\s_X\+(t) \s_X(t)\> \<\s_X\+(t+\tau) \s_X(t+\tau)\>\nn
&-\big|\<\s_X\+(t+\tau)\s_X(t)\>\big|^2
+G^{(2)}(t,\tau)\big]\, ,
\end{align}
\end{subequations}
where the last term in Eq.~\eqref{eq:G2HOM_t_tau} accounts for nonunity single-photon purities. This quantity is measured in a Hong-Ou-Mandel setup \cite{Hong1987}. Perfect indistinguishability corresponds to $\I=1$ and using the definition Eq.~\eqref{eq:G2HOM_t_tau} it is bounded by $0.5\leq\I\leq1$ \cite{Fischer2016}. We note that other definitions of $\I$ are often used which are not applicable when the single-photon purity deviates from unity and where the lower bound is $0$ rather than $0.5$ \cite{Manson2016,Iles-Smith2017b}.

%\textbf{Brightness}
The brightness $\B$ of a photon source is defined as the number of photons emitted per excitation laser pulse \cite{Somaschi2016}. It is given as \cite{Gustin2018,Cosacchi2019}
\begin{align}
\label{eq:B}
\mathcal{B}:=\,&\gamma \int_{t_0-T_{\t{Pulse}}/2}^{t_0+T_{\t{Pulse}}/2} dt\, \<\s_X\+(t) \s_X(t)\>\, ,
\end{align}
where $t_0$ is the center time of the pulse and $0\leq\mathcal{B}\leq\g T_{\t{Pulse}}$.
We scale $\B$ such that $100\,\%$ corresponds
to the ideal case of a delta-pulse excitation.

To calculate these quantities we use the path-integral method both without and with the QRT. 
%%% Expand on PI without QRT
The path-integral method propagates the augmented density matrix that contains the information about the memory induced by the environment to the QD dynamics.
Since the phonon-induced memory depth is finite, a memory window is formed in each time step.
To implement the QRT, 
%instead of transporting the memory window over from the $t$- to the $\tau$-propagation,
the augmented density matrix is traced over all memory-related variables at the end of the $t$-propagation to yield a new initial reduced density matrix before the subsequent $\tau$-propagation. Thus, the accumulated phonon memory is discarded for the $\tau$-propagation.
Therefore, the $\tau$-propagation becomes independent from the past evolution in $t$, which is the central assumption of the QRT. We have checked the validity of this approach by comparing our results with a standard implementation of the QRT as discussed in Ref.~\onlinecite{McCutcheon2016} and verify the finding therein that the QRT yields the phonon sidebands in emission spectra on the energetically wrong side, cf., Fig.~2 in the SM \cite{supp}.

%\textbf{Results1} 
Using the path-integral method, we calculate the photon properties $\P$, $\I$, and $\B$ for a wide parameter range as shown in Fig.~\ref{fig:I_and_B}, which displays the results using the path-integral approach without the QRT approximation. In the phonon-free case, $\lambda=0$, the excitation of the QD leads to a near-optimal single-photon source with $\P=99.76\,\%$, $\I=99.76\,\%$, and $\B=99.82\,\%$. Slight deviations ($<0.3\,\%$) from the perfect source can be traced back to the finite pulse length.

While the single-photon purity is close to unity for the entire parameter range under consideration, for a finite phonon scaling $\lambda$, the indistinguishability rapidly deteriorates with rising temperature $T$, such that for $\lambda=1$ it falls below $70\,\%$ when $T>30\,$K. For large phonon scalings, the indistinguishability cannot exceed $60\,\%$ even at $T=4\,$K. At higher temperatures and for large phonon scaling, the indistinguishability decreases to its lowest possible value of $50\,\%$. Nonetheless, the corresponding brightness is nonvanishing, such that the QD becomes a source of distinguishable single photons in this regime of higher temperatures and stronger QD-phonon coupling.

We have marked the physically most relevant regions with yellow boxes in Fig.~\ref{fig:I_and_B}(c). They correspond to the low-temperature regime in which experiments are typically conducted for different QD materials from GaAs to GaN modeled here by different scalings $\lambda$ (vertical box) as well as over a temperature range between liquid helium and nitrogen temperatures (horizontal box) for GaAs ($\lambda=1$). 
In the parameter range of highest interest, i.e., where the boxes overlap at $\lambda=1$ and $T=4\,$K, we find $\P=99.79\,\%$, $\I=93.16\,\%$, and $\B=96.75\,\%$.

%\textbf{Non-Markovianity measure:}
We now evaluate how the QRT approximation changes these results.
It is usually conjectured that the QRT might fail when the dynamics is non-Markovian, i.e.,  when memory effects are non-negligible \cite{Breuer2016,deVega2017}. Furthermore, there is a class of environmental couplings for which the QRT cannot be accurately applied, even when the single-time dynamics is Markovian \cite{Guarnieri2014}.
In order to describe the contribution of the memory effects quantitatively, we consider a non-Markovianity measure for our system.  

In contrast to classical Markovian stochastic processes, in open quantum systems there is no single definition of Markovianity (or non-Markovianity) that is agreed upon. Rather, there are different measures that capture different aspects of Markovian quantum dynamics \cite{Breuer2009,Rivas2010,Lu2010,Lorenzo2013, Rivas2014, Chruscinski2014,Li2018}, one of which is the trace distance measure. The trace distance between two states described by the reduced density matrices $\r_1$ and $\r_2$ is defined as
\begin{align}
D(\r_1(t),\r_2(t)):=\frac{1}{2}||\r_1(t)-\r_2(t)||_1=\frac{1}{2}\sum_k|x_k(t)|\, ,
\end{align}
where $x_k(t)$ are the eigenvalues of the difference matrix $\r_1(t)-\r_2(t)$. In our case, $\r_1$ and $\r_2$ correspond to arbitrary states chosen on the Bloch sphere of the two-level QD.

For Markovian dynamics, this quantity is a contraction
\begin{align}
\frac{d}{dt}D(\r_1(t),\r_2(t))\leq0\, .
\end{align}
The intuitive explanation for this behavior lies in the loss of information in a Markovian system: two originally distinct states monotonically lose their distinguishability  over time.
Only in a non-Markovian system, information can flow back from the environment to the system, making the trace distance a non-monotonic function of time.
Therefore, the non-Markovianity of a system can be quantified as \cite{Breuer2009,Guarnieri2014,McCutcheon2016}
\begin{align}
\label{eq:N}
\N:=\max_{\r_1,\r_2}\int_{\Omega_+}\,\frac{d}{dt}D(\r_1(t),\r_2(t))\,dt\, .
\end{align}
$\Omega_+$ is the union of the intervals on which $\frac{d}{dt}D(\r_1(t),\r_2(t))>0$.
The maximum is taken over all pairs of possible initial states.
Fortunately, only the subset of states needs to be considered that are orthogonal to each other \cite{Wissmann2012}. For our two-level system, this means that the corresponding Bloch sphere needs to be sampled only for pairs of opposing points on its surface.
%Moreover, only one hemisphere has to be considered since the trace distance is symmetric in its two arguments. We numerically evaluate Eq.~\eqref{eq:N} by discretizing the two angles describing the surface of the Bloch hemisphere to yield a net of 1024 points.

While $\N=0$ implies Markovianity, it is important to realize that $\N\neq0$ implies that the underlying dynamical map is \textit{indivisible} \cite{Guarnieri2014}. Therefore, the measure $\N$ captures the appearance of memory effects in the dynamics of the system, which is a fundamental aspect of non-Markovianity both in classical stochastic processes and open quantum systems.

\begin{figure*}[t]
	\centering
	\includegraphics[width=\textwidth]{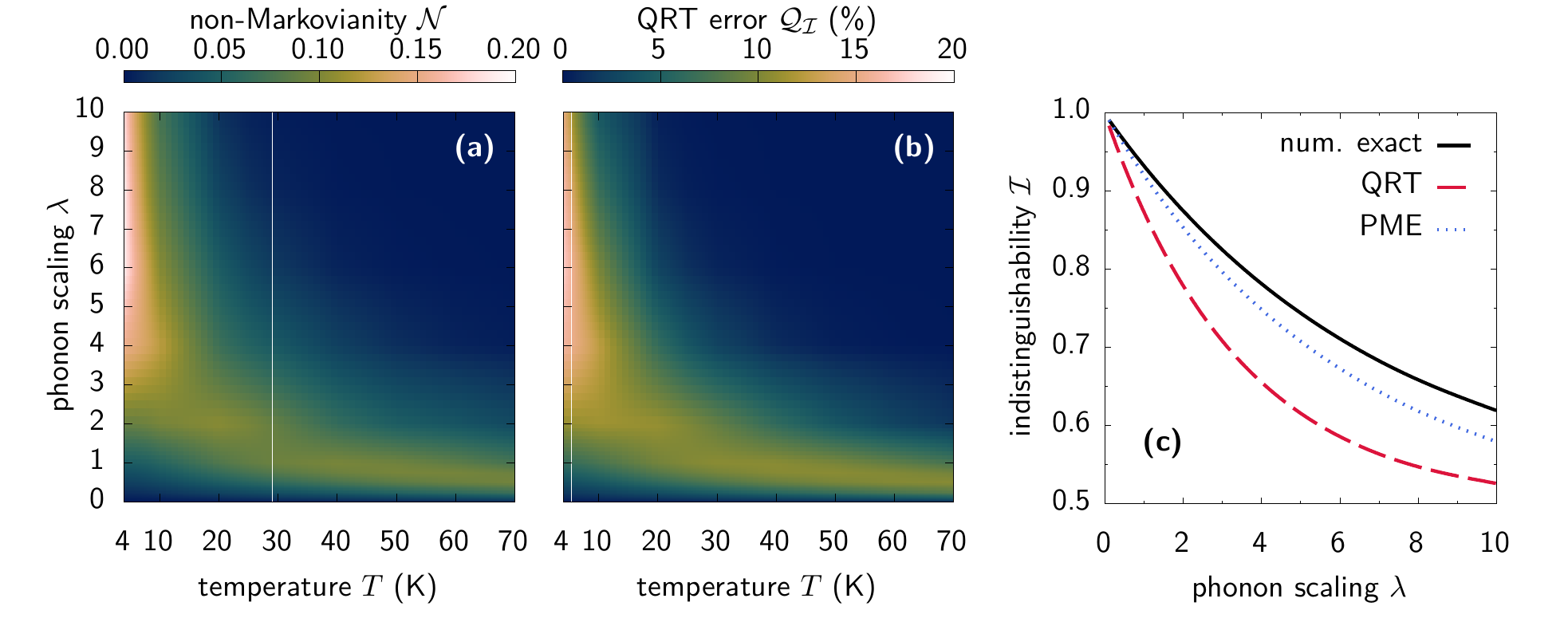}
	\caption{The non-Markovianity measure $\N$ (a) and the relative error $\Q_\I$ for the indistinguishability (b) as a function of temperature $T$ and phonon scaling $\lambda$.
	(c) The indistinguishability as a function of the phonon scaling parameter $\l$ at $4\,$K, calculated with the numerically exact path-integral method (num. exact), by using the QRT in the lab frame (QRT), and by applying the QRT in the polaron transformed frame within the PME approach (PME). }
	\label{fig:N_and_QI}
\end{figure*}

%\textbf{Testing the QRT}
To quantify the deviations introduced by the QRT, we define the relative error of evaluating a target quantity $M$ using the QRT as a measure for the validity of the QRT with respect to $M$:
\begin{align}
\Q_M=\left|\frac{M-M_{\t{QRT}}}{M}\right|\, ,
\end{align}
where $M$ is calculated numerically exact and $M_{\t{QRT}}$ using the QRT.
 
The QRT states that the same dynamical map that is used to evolve the density matrix and, in extension, expectation values of any subsystem operator, can be used for the time evolution of multitime correlation functions used in Eq.~\eqref{eq:G2_t_tau} and \eqref{eq:G2HOM_t_tau}. In particular, the differential equation propagating the density matrix in the real time $t$ is reused for the propagation in the delay time $\tau$ \cite{Louisell1973,Carmichael1993}. This assumption presumes that the initial factorization of subsystem and environment common in the description of open quantum systems is also used at the beginning of the $\tau$-dynamics. In other words, this factorization is assumed for every $t$.

%\textbf{Results}
Now, we examine the impact of the QRT approximation on the photon source characteristics considered above. The non-Markovianity measure $\N$ and the relative error $\Q_\I$ for the indistinguishability are depicted in Fig.~\ref{fig:N_and_QI}(a) and (b) as a function of $T$ and $\lambda$. We see large values of $\N$ and $\Q_\I$, in particular, in the physically relevant parameter regimes, i.e., at $\lambda=1$ and low temperatures. The largest $\N$ is found for $\lambda>1$ and $T<10\,$K [cf., Fig.~\ref{fig:N_and_QI}(a)], where also the error introduced by using the QRT rises up to roughly $18\,\%$. This behavior can be related to the connection between Markovianity and the QRT. Interestingly, there are also parameter ranges with a nonzero $\N$, where the QRT error is insignificant, e.g., at $\lambda=10$ and $T=20\,$K, where $\N=0.0125$, while $\Q_\I=0.3\,\%$.
This means that there are parameter sets where the QRT approximation is valid to a better degree than a Markovian description.
This is unexpected since the former imposes more restrictive conditions on the system dynamics: for the QRT to hold, the subsystem and environment have to factorize for all times $t$, not only at the initial time.
In the entire parameter regime considered here, the QRT overestimates the phonon influence on $\I$, that is $\I>\I_{\t{QRT}}$, cf., Fig.~\ref{fig:N_and_QI}(c) for a slice at $4\,$K. 

In contrast, the error $\Q_\P$ introduced by the QRT to the single-photon purity is negligible and the brightness is unaffected by the QRT, since its definition in Eq.~\eqref{eq:B} contains only expectation values at a single time.
Surprisingly, $\Q_\P$ is also extraordinarily small, being on the order of $10^{-4}$ for all considered parameter values (not shown), in contrast to $\Q_\I$.

In order to understand this, we examine the multitime correlation functions. While the purity contains only the second-order correlation $G^{(2)}(t,\tau)$, the indistinguishability also includes the correlation $G^{(1)}(t,\tau):=\<\s_X\+(t+\tau)\s_X(t)\>$. In $G^{(2)}(t,\tau)$ the operators $\s_X\+$ and $\s_X$ appear in pairs at each time $t$ and $t+\tau$, respectively, hence modeling intensity-intensity correlation measurements, i.e., the correlation between occupations. In $G^{(1)}(t,\tau)$ on the other hand, $\s_X\+$ and $\s_X$ appear as standalone operators for each time argument in $G^{(1)}(t,\tau)$. Therefore, this function correlates coherences rather than occupations. Because the coupling to the LA phonon environment has a stronger impact on coherences than on occupations, it becomes clear why the approximations introduced by the QRT have a significantly stronger impact on $\I$ than on $\P$.

This finding implies two consequences: first, the single-photon purity can be calculated using the QRT with negligible error, even for those parameters, where the dynamics is clearly non-Markovian according to the measure $\N$ [cf., Fig.~\ref{fig:N_and_QI}(a)]. Second, one cannot use $\Q_\P$ as a \textit{general} measure for the validity of the QRT. Using it in such a way would imply the validity of the QRT, which is misleading since in the same parameter regimes considered, the indistinguishability is off by up to $18\,\%$ when evaluating with the QRT.

%Moreover, as described in detail in the SM \cite{supp}, we have checked the QRT's success in a polaron transformed frame. Indeed, changing the frame improves the usage of the QRT, but still a significant error in the photon indistinguishability is obtained.

Finally, we analyze the frame dependence of the QRT by applying it in a polaron transformed frame.
This technique is widely used in the polaron master equation approach (PME) \cite{Roy2011b,Nazir2016,Iles-Smith2016}, see also SM \cite{supp}.
In Fig.~\ref{fig:N_and_QI}(c), the indistinguishability is shown for varying phonon scaling parameter $\l$ at $T=4\,$K.
The numerically exact result (black solid line) is compared with the calculation using the QRT in the lab frame (red dashed line) and the PME approach applying the QRT in the polaron frame (blue dotted line).
While all methods yield qualitatively the same dependency, the PME produces results closer to the numerically exact calculation.
While the largest relative error encountered in the slice shown in Fig.~\ref{fig:N_and_QI}(c) is $18\,\%$ for the QRT in the lab frame (red dashed line), it is only $6\,\%$ when the QRT is applied in the polaron frame within the PME.
The better performance of the PME is expected because due to the transformation to the polaron frame a variety of, but not all non-Markovian effects are captured.
Therefore, changing the frame improves the usage of the QRT, but still a significant systematic overestimation of phonon effects on the photon indistinguishability is obtained.

In summary, assessing the validity of the commonly used QRT is dependent on the target quantity that is calculated.
In particular, there is no single measure by which the validity of the QRT could be estimated for all possible figures of merit derived from multitime correlation functions.
Using a numerically exact path-integral method to calculate the properties of photons emitted from a QD coupled to LA phonons enabled us to explore the boundaries of the QRT, showing that the phonon effect on photon indistinguishability is systematically overestimated by the QRT, while the purity can be safely calculated using the QRT.
Unlike what is found for other systems \cite{Guarnieri2014}, the QRT induces errors in the photon emission from QDs typically only when the dynamics is non-Markovian.
Though we show that due to the phonons the photon properties are limited close to but below unity in typical cases, there is still room for improvement, e.g., by placing the QD in a cavity.
Furthermore, our results should be applicable to a broad range of physical two-level systems, such as defects in diamonds \cite{Englund2010,Beha2012,Aharonovich2016,Fehler2019,
Janitz2020,Schrinner2020}, silicon \cite{Lee2012,Fehler2020}, hexagonal boron nitride \cite{Proscia2020,Froech2020}, or other solid-state emitters \cite{Rodt2020} coupled to a continuum of environmental oscillators.

\acknowledgments
This work was funded by the Deutsche Forschungsgemeinschaft (DFG, German Research Foundation) - project Nr. 419036043.

\bibliographystyle{appa}
\bibliography{bib,new}
\end{document}